\documentclass[twocolumn,aps,prc,preprintnumbers,showpacs,showkeys,nofootinbib,superscriptaddress,fleqn,floatfix,tightenlines, 10pt]{revtex4-1}
\usepackage{amsmath,amsfonts,amssymb,amscd,amsxtra,amsthm}
\usepackage{natbib}
\usepackage{graphicx,subfigure}  
\usepackage{epstopdf}
\usepackage{dcolumn}  
\usepackage{bm}          
\usepackage{slashed}
\usepackage[utf8]{inputenc}
\usepackage{CJK}
\usepackage{mathptmx}
\usepackage{mathrsfs}
\usepackage[normalem]{ulem} 
\usepackage[dvipsnames]{xcolor} 
\usepackage{array}
\usepackage{multirow}

\def\vc#1{\mbox{\boldmath $#1$}}

\begin{document}
\title{Cluster structures of the $0^+$ states of $^{13}_\Lambda{\rm C}$ with the 3$\alpha+\Lambda$ four-body orthogonality condition model}

\author{Qian Wu}
\email[E-mail: ]{qianwu@impcas.ac.cn}
\affiliation{Institute of Modern Physics, Chinese Academy of Science, Lanzhou 730000, China}

\author{Yasuro Funaki}
\email[E-mail: ]{yasuro@kanto-gakuin.ac.jp}
\affiliation{College of Science and Engineering, Kanto Gakuin University,
Yokohama 236-8501, Japan}
\affiliation{RIKEN Nishina Center,RIKEN,2-1 Hirosawa,351-0115
  Saitama,Japan}

\author{Xurong Chen}
\affiliation{Institute of Modern Physics, Chinese Academy of Sciences, Lanzhou 730000, China}
\affiliation{School of Nuclear Science and Technology, University of Chinese Academy of Sciences, Beijing 100049, China}
\affiliation{Guangdong Provincial Key Laboratory of Nuclear Science, Institute of Quantum Matter,
South China Normal University, Guangzhou 510006, China}

\begin{abstract}
We study structures of the $0^+$ states of $^{13}_\Lambda$C within the framework of $\alpha+\alpha+\alpha+\Lambda$ four-body orthogonality condition model with the Gaussian expansion method. We obtain five $0^+$ states of $^{13}_\Lambda$C and compare them with the low-lying four $0^+$ states of $^{12}$C. We find that the $0_1^+$, $0_2^+$ and $0_4^+$ states correspond to the $0_1^+$, $0_2^+$ and $0_4^+$ states of $^{12}$C, respectively. We also find that both the $0_3^+$ and $0_5^+$ states have analogous nature to the $0_3^+$ state of $^{12}$C, where the additional $\Lambda$ particle couples with $^8$Be cluster and one $\alpha$ cluster, in the $0_3^+$ state of $^{12}$C, respectively. The former is then shown to have an elongated $^9_\Lambda$Be$(0_1^+) + \alpha$ cluster structure with a higher nodal behaviour for their relative wave function, and the latter have a dilute three-body gas consisting of $2\alpha+^5_\Lambda$He clusters.
\keywords{OCM, hypernuclei, Microscopic cluster model}
\end{abstract}
\maketitle
\section{Introduction}
Nuclear clustering is extremely important in understanding the structures of nuclear excitation~\cite{wildermuth}. According to the threshold rule, the so-called Ikeda diagram \cite{ikedagram} tells us that the cluster structure appears around the corresponding clusters breaking up threshold. The $0^+_2$ at 7.65 MeV in $^{12}$C, known as the Hoyle state, is a typical example of the cluster states. It is predicted by Fred Holye \cite{Hoyle1954} at 1954 which is a gateway state for massive $^{12}$C production in stellar evolution and subsequently observed by Cook et al \cite{Cook1957} at 1957. There is a long history in studying the structure of the Hoyle state. In the early stage, Morinaga proposed a linear chain structure of the $3\alpha$ clusters for the Hoyle state \cite{Morinag1956}. However in 1975, with the semi-microscopic cluster model, known as the Orthogonality Condition Model (OCM), Horiuchi proposed that the Hoyle state has a weakly coupled $3\alpha$-cluster structure with relative $S$-waves~\cite{HoriuchiOCM}. That was soon confirmed by Kamimura~\cite{kamimura1981RGM} and Uegaki~\cite{Uegaki1977} with the fully microscopic cluster model calculations, Resonating Group Method (RGM) and the Generator Coordinate Method (GCM), respectively. After a few decades, an interesting conjecture was proposed that the Hoyle state has a Bose-condensate-like structure, where all $3\alpha$ particles weakly interact with each other and occupy an identical $S$-orbit~\cite{THSR2001}. This new interpretation was well established by the fact that the so-called Tohsaki-Horiuchi-Schuck-R\"opke (THSR) wave function was proved to be almost equal to the RGM or GCM wave function for the Holyle state~\cite{Funaki2003,Funaki2002}.


Moreover, the observed $0^+$ state at 10.3 MeV has been known for a long time as a resonance with a broad width of about 3 MeV. Recently, Itoh et al.\cite{Itoh_2013} reported that this broad $0^+$ state is decomposed into two states, the $0^+_3$ and $0^+_4$ states at 9.04 MeV and 10.56 MeV, with the width of 1.45 MeV and 1.42 MeV, respectively. These resonance parameters are well reproduced by the OCM $+$ Complex Scaling Method (CSM) calculations~\cite{OCM2007,kami2013}. However, in several calculations without imposing a resonance boundary condition, such like the Antisymmetrized Molecular Dynamics (AMD)~\cite{AMD2007}, the Fermionic Molecular Dynamics (FMD)~\cite{FMD2007} and the GCM~\cite{GCM2015}, it seems that the observed new $0_3^+$ state is missing and only the new $0^+_4$ state is reproduced. As for the structures of the $0^+_3$ state, the OCM calculation in Ref.\cite{OCM2007} claimed that it has an $S$-wave higher nodal structure for the relative wave function between the $^8$Be and $\alpha$ clusters. Although the observed $0^+_3$ state is missing in the AMD and FMD calculations~\cite{AMD2007,FMD2007}, they give a single $0^+$ state above the Hoyle state, which has a linear-chain-like structure of the $3\alpha$ clusters and may correspond to the observed $0^+_4$ state. Later this linear-chain-like structure for the $0_4^+$ state was confirmed with the THSR wave function, together with the simultaneous description of the $0_3^+$ state, which behaves like an excited state of the Hoyle state~\cite{funaki2016}.

While the Hoyle state, which has a well-developed $3\alpha$ cluster structure, locates slightly (0.38 MeV) above the $3\alpha$ breakup threshold, in $^{13}_\Lambda$C, the other new thresholds decaying into the $^{9}_\Lambda{\rm Be}+\alpha$, $^{5}_\Lambda{\rm He}+^8{\rm Be}$ and $^{12}{\rm C}+\Lambda$ clusters appear below the $3\alpha+\Lambda$ four-body breakup threshold (see Fig.~\ref{fig:0pspec}). Since the cluster states are likely to appear around the corresponding thresholds, according to the threshold rule, it seems that more variety of cluster structures appear and hence the clustering becomes more important, when a $\Lambda$ particle is added into a core nucleus, i.e. like in $^{13}_\Lambda$C.

Concerning the single $\Lambda$ hypernucleus, the simple $2\alpha+\Lambda$ system, $^{9}_\Lambda$Be, has already been well established by many cluster model calculations~\cite{qian2020,Lee2019few,Hiyama1997ptep,AMDbe9l,Bando1983ptp,Motoba1985ptp,Yamada1988prc}. $^{13}_\Lambda$C has also been investigated by using the microscopic and semi-microscopic cluster models, RGM, GCM, and OCM. In Ref.\cite{Hiyama1997ptep}, Hiyama~{\it et al.} fully solved the $3\alpha + \Lambda$ four-body problem in $^{13}_\Lambda$C with the OCM and the Gaussian expansion method~\cite{Hiyama2003gem}. They obtained the $(1/2)_1^+$ and $(1/2)_2^+$ states and concluded that both the states correspond to the $0^+_1$  and the Hoyle state ($0_2^+$) in $^{12}$C, respectively, where the former state has almost no change of the spatial size of the $^{12}{\rm C}$ core and the latter has a drastic spatial shrinkage from the $^{12}$C core, by adding the $\Lambda$ particle. Yamada~{\it et al.} studied the structures of the $(1/2)_1^+$, $(1/2)_2^+$, and $(1/2)_3^+$ states by solving the $\Lambda + {^{12}{\rm C}}$ coupled channel equation~\cite{Yamada1985ptep}, where $^{12}{\rm C}$ is described by the $3\alpha$ RGM wave function~\cite{kamimura1981RGM}. They found that the $(1/2)_1^+$, $(1/2)_2^+$ and $(1/2)_3^+$ states are dominated by the components of $\Lambda(0s)+^{12}{\rm C}(0_1^+)$, $\Lambda(0s)+^{12}{\rm C}(0_2^+)$, and $\Lambda(1s)+^{12}{\rm C}(0_1^+)$ configurations, respectively. However, all these calculations of $^{13}_\Lambda{\rm C}$ were done within the bound state approximation, and also higher excitation energy region up to around the $3\alpha + \Lambda$ threshold has never been investigated.

The study of exploring the higher excitation energy region in $^{13}_\Lambda$C was, however, done by using the so-called Hyper-THSR wave function (H-THSR), where the THSR wave function is adapted to apply to the $\Lambda$ hypernucleus~\cite{Funaki2014ptep,Funaki2018plb}, though any correct boundary condition of resonances is not imposed. The authors obtained the four $0^+$ states, neglecting the spin-orbit coupling between the $\Lambda$ particle and core, and compared them with the $0_1^+$-$0_4^+$ states of $^{12}{\rm C}$. While the $0_1^+$ and $0_2^+$ states of $^{13}_\Lambda$C are considered to correspond to the states, in which the $\Lambda$ particle couples to the ground state and the Hoyle state in $^{12}{\rm C}$, respectively, the $0_3^+$ and $0_4^+$ states are considered to correspond to the $0_4^+$ and $0_3^+$ states in $^{12}{\rm C}$, respectively. It was shown that the former has the linear-chain-like structure of the $3\alpha$ clusters with the $\Lambda$ particle and the latter has the more dilute gaslike structure of $^5_\Lambda{\rm He}+2\alpha$ clusters, as being the analog to the Hoyle state. It was also argued that the $\Lambda$ particle only couples to one $\alpha$ cluster to form $^5_\Lambda{\rm He}$ in the $0_4^+$ state, and therefore the state is less bound than the $0_3^+$ state, in which the additional $\Lambda$ paticle couples all the $\alpha$ clusters, keeping the $3\alpha$ linear-chain-like structure.


In this work, we study the structures of the $0^+$ states of $^{13}_\Lambda$C up to around the $3\alpha+\Lambda$ breakup threshold within the $3\alpha + \Lambda$ four-body OCM, where the relative motions between the clusters are fully solved by using the Gaussian basis functions. Relative angular momentum channels are taken to cover all the possible configurations including the $^{12}$C+$\Lambda$, $^8$Be+$^5_\Lambda$He and $^9_\Lambda$Be+$\alpha$ clustering in $^{13}_\Lambda$C. We use the $\alpha$-$\alpha$ and $\Lambda$N interaction, which reproduce the binding energies of $^8$Be, $^5_\Lambda$He, the ground and the first excited states of $^9_\Lambda$Be, and the ground state of $^{13}_\Lambda$C.
We obtain five $0^+$ states in $^{13}_\Lambda$C. With calculating the S$^2$ factors and mass radius, we find that the $0_1^+$, $0_2^+$ and $0_4^+$ states in $^{13}_\Lambda$C correspond to the $0_1^+$, $0_2^+$ and $0_4^+$ states of $^{12}$C, respectively. Both the $0_3^+$ and $0_5^+$ states have an analogous structure to the $0_3^+$ state of $^{12}$C, where the additional $\Lambda$ particle couples with $^8$Be cluster and one $\alpha$ cluster, in the $0_3^+$ state of $^{12}$C, respectively.

The paper is organized as follows. In Sec.~\ref{sec:II}, we introduce the realistic NN and $\Lambda$N interaction and the unique adjustments of the potential parameters. The energy levels of the $0_1^+$-$0_5^+$ states and their rms radii are discussed in Sec.~\ref{sec:IIIa}. In Sec.~\ref{sec:IIIb} and Sec.~\ref{sec:IIIc}, the Reduced Width Amplitudes (RWAs) and the $S^2$-factors of the $0_1^+$, $0_2^+$ and $0_4^+$ states and the $0_3^+$ and $0_5^+$ states are discussed, respectively, and their structures are compared with those of the corresponding $0^+$ state of $^{12}{\rm C}$. Sec.~\ref{sec:IV} is devoted to Summary.

\section{Hamiltonian and Method}\label{sec:II}
We solve a four-body Schr\"odinger equation in $^{13}_{\Lambda}{\rm C}$ composed of the $3\alpha+\Lambda$ clusters with the semi-microscopic cluster model, the OCM, where the hamiltonian is then written as follows:
\begin{equation}
H=T+\sum_{i<j=1}^{3}V_{\alpha_{i} \alpha_{j}}+\sum_{i=1}^{3} V_{\alpha_{i} \Lambda}+V_{\alpha_1\alpha_2\alpha_3}+\sum_{i<j=1}^{3}V_{\alpha_{i} \alpha_{j}}^{\text {Pauli }}, \label{eq:hamil}
\end{equation}
where $T$ is the kinetic energy operator for the relative motions between the clusters, and the $V_{\alpha_{i} \alpha_{j}}$ and $V_{\alpha_{i} \Lambda}$ are the $\alpha-\alpha$ and $\alpha-\Lambda$ interaction operators, respectively. The $V_{\alpha_{i} \alpha_{j}}^{\text{Pauli}}$ and $V_{\alpha_1\alpha_2\alpha_3}$ represent the Pauli exclusion operator between the $\alpha$ clusters and the phenomenological 3-body interaction operator between the $3\alpha$ clusters, respectively, which will be explained later in detail.

\begin{figure}[htbp]
\setlength{\abovecaptionskip}{0.cm}
\setlength{\belowcaptionskip}{-0.cm}
\centering
\includegraphics[width=0.45\textwidth]{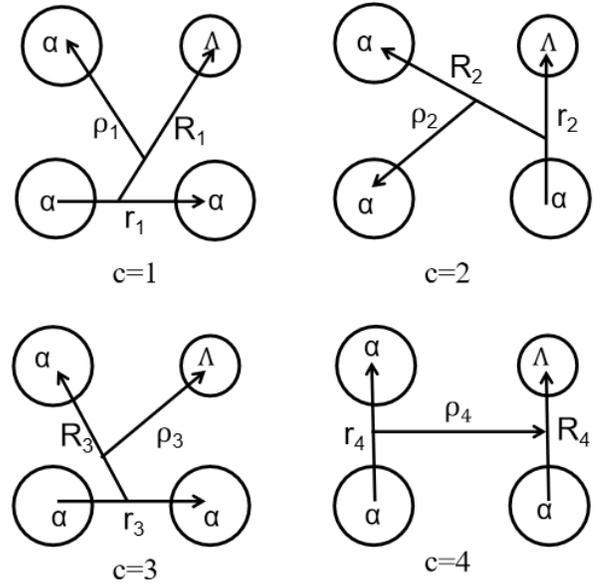}
\caption{
Jacobi coordinates of $^{13}_{\Lambda}$C with $3\alpha +\Lambda$ model.}
\label{fig:c13jacobi}
\end{figure}

The total wave function with the $J^\pi=0^+$ of the four-body $3\alpha+\Lambda$ cluster model can be described as follows:
\begin{equation}
\begin{aligned}
\Psi \left(_{\Lambda}^{13} \mathrm{C}\right)&= \sum_{c=1}^{4} \sum_{L} \sum_{n_1,l_1}\sum_{n_2,l_2}\sum_{n_3,l_3} C_{
n_1 l_1 n_2 l_2 n_3 l_3 L }^{(c)} \\
& \times \mathcal{A}_{\alpha}\Big[ \Phi^{\left(\alpha_{1}\right)} \Phi^{\left(\alpha_{2}\right)} \Phi^{\left(\alpha_{3}\right)} \\
& \times \left[\left[\phi_{n_1 l_1}^{(c)}\left(\boldsymbol{r}_{c}\right) \psi_{n_2 l_2}^{(\mathrm{c})}\left(\boldsymbol{R}_{c}\right)\right]_{L} \varphi_{n_3 l_3}^{(c)}\left(\boldsymbol{\rho}_{c}\right)\right]_{J=0}\Big] , \label{eq:totalwf}
\end{aligned}
\end{equation}
where the four sets of Jacobian coordinates ($c=1-4$), shown in Fig.~\ref{fig:c13jacobi}, are used, the $\mathcal{A}_{\alpha}$ is the $\alpha-\alpha$ symmetrization operator, and $\Phi^{(\alpha_{i})}$ with $i=1,2,3$ are the intrinsic wave functions of the $\alpha$ cluster with the $(0s)^4$ configuration.
The relative wave functions between the clusters, corresponding to the three Jacobi coordinates, $\phi_{n_1 l_1}(\boldsymbol{r})$, $\psi_{n_2 l_2}(\boldsymbol{R})$ and $\varphi_{n_3 l_3}(\boldsymbol{\rho})$, are expanded by using the following Gaussian basis functions, applying the Gaussian Expansion Method~\cite{Hiyama2003gem,Hiyama2009ppnp},
\begin{eqnarray}
&\phi_{n_1 l_1}(\boldsymbol{r})=r^{\ell_1} e^{-(r/r_{n_1})^2}Y_{\ell_1 m_1}(\hat{\vc r}), \nonumber \\
&\psi_{n_2 l_2}(\boldsymbol{R})=R^{\ell_2} e^{-(R/R_{n_2})^2}Y_{\ell_2 m_2}(\hat{\vc R}), \nonumber \\
&\varphi_{n_3 l_3}(\boldsymbol{\rho})=\rho^{\ell_3} e^{-(\rho/\rho_{n_3})^2}Y_{\ell_3 m_3}(\hat{\vc \rho}).
\end{eqnarray}
The Gaussian variational parameters are chosen to have geometric progression below,
\begin{eqnarray}
&r_{n_1}=r_{\rm min} A_1^{n_1-1},  \quad (n_1=1 \sim n_1^{\rm max}), \nonumber \\
&R_{n_2}=R_{\rm min} A_2^{n_2-1},  \quad (n_2=1 \sim n_2^{\rm max}), \nonumber \\
&\rho_{n_3}=\rho_{\rm min} A_3^{n_3-1},  \quad (n_3=1 \sim n_3^{\rm max}).
\end{eqnarray}
Since the energy splitting of the $3/2^+-5/2^+$ is small enough only to be measured by the high resolution $\gamma$-ray experiments\cite{Hiyama2009ppnp,Akikawa2002be9lgamma,Tamura2004gammaray}, we neglect the spin orbit coupling between the $\Lambda$ and $\alpha$ clusters.
The eigen energies and the coefficients $C_{n_1 l_1 n_2 l_2 n_3 l_3 L}^{(c)}$ are obtained with applying the Rayleigh-Ritz variational method.

The $\alpha$-$\alpha$ interaction is constructed by the folding procedure from the modified Hasegawa-Nagata (MHN) effective nucleon-nucleon potential~\cite{hasegawa1971aa} and proton-proton coulomb potential, so as to reproduce the observed $\alpha$-$\alpha$ scattering phase shift and the binding energy of the ground state of $^8$Be within the $\alpha$-$\alpha$ OCM. The Pauli principle between the two $\alpha$ clusters in the OCM is taken into account by introducing the following Pauli exclusion operator into the Hamiltonian in Eq.~(\ref{eq:hamil}),
\begin{equation}
V_{\alpha_{i} \alpha_{j}}^{\text {Pauli }} = \lim\limits_{\lambda \rightarrow \infty}\lambda\sum_{f=0s,1s,0d}
|\phi_f ({\vc r}_{\alpha_i \alpha_j})\rangle \langle \phi_f ({\vc r}^\prime_{\alpha_i \alpha_j})|.
\end{equation}
The Pauli forbidden states($0s,1s,0d$) are ruled out when $\lambda$ is an infinity
and practically the $\lambda$ is given around $\sim10^5$MeV, which is high enough to push
the nonphysical states into a large energy region without affecting the physical states.
The harmonic oscillator size parameter of the $0s$ wave function of the $\alpha$ cluster, $\Phi^{(\alpha_{i})}$ with $i=1,2,3$ in Eq.~(\ref{eq:totalwf}), is taken as $1.358$ fm.

It is, however, well known that the folding $\alpha$-$\alpha$ potential given by the MHN force significantly overbinds $^{12}$C, as discussed in Ref.\cite{Hiyama1997ptep}. We therefore introduce the following phenomenological three-body repulsive force between the $3\alpha$ clusters,
\begin{equation}
V_{\alpha_1\alpha_2\alpha_3}=V_{0} \exp \left[-\mu\left(r_{\alpha_{1} \alpha_{2}}^{2}+r_{\alpha_{2} \alpha_{3}}^{2}+r_{\alpha_{3} \alpha_{1}}^{2}\right)\right],
\end{equation}
where $V_0=154$ MeV and $\mu=0.195\ {\rm fm}^{-2}$ are adopted, to reproduce the observed binding energy of the ground state of $^{12}$C, $E=-7.27$ MeV, relative to the $3\alpha$ threshold. Then the Hoyle state is located at $E=0.86$ MeV above the $3\alpha$ threshold, while the experimental value is $0.38$ MeV.

The $\Lambda$-$\alpha$ interaction is also obtained by the folding procedure of the $\Lambda$N interaction. We adopt the so-called YNG interaction~\cite{Yamamoto1994yng} as the $\Lambda$N interaction, which simulates the G-matrix $\Lambda$N interaction derived from the Nijmegen model f (NF), of the three-range Gaussian form as a function of fermi momentum $k_F$, given by,
\begin{eqnarray}
V_{\Lambda N}(r,k_F)=\sum\limits_{i=1}^{3}[(v_{0,even}^i+v_{\sigma\sigma,even}^i\sigma_{\Lambda}\cdot\sigma_N)\frac{1+P_r}{2} \nonumber \\
+(v_{0,odd}^i+v_{\sigma\sigma,odd}^i\sigma_{\Lambda}\cdot\sigma_N)\frac{1-P_r}{2}]e^{-(r/\beta_i)^2},
\end{eqnarray}
where $P_r$ is the space exchange operator. However, it should be noted that with this NF interaction, the ground state of $^9_\Lambda$Be is overbound, due to the strong attraction of odd-state component of spin independent part of the $\Lambda N$ interaction \cite{Hiyama1997ptep}. We thus tune the odd state part of the $\Lambda N$ interaction to reproduce the observed binding energy of $^5_\Lambda$He and $^9_\Lambda$Be. The parameters of this modified NF interaction are listed in Table \ref{tab:YNG}.
\begin{table}[tbh]
\caption{Parameters of the modified YNG-NF $\Lambda N$ interaction, where $k_F=0.963$ fm$^{-1}$ is taken.
}
\begin{tabular}{p{2cm}<{\centering}p{1.5cm}<{\centering}
p{1.5cm}<{\centering}p{1.5cm}<{\centering}}
  \hline\hline
 $\beta_i$ (fm)& 1.50 &0.90 &0.50         \\
  \hline
 $v_{0,{\rm even}}^i$ & $-9.22$ &$-187.63$&795.43  \\
 $v_{0,{\rm odd}}^i$ & $-5.67$&$-35.26$&2141.79(1172.21)   \\
 \hline
\end{tabular}
\label{tab:YNG}
\end{table}

Then, with the $\alpha-\Lambda$ and $\alpha-\alpha$ interaction mentioned above, the subsystems of $^{13}_\Lambda$C, including $^8$Be($0^+$), $^8$Be($2^+$), $^5_\Lambda$He, $^{12}$C($0_1^+$), $^{12}$C($0_2^+$), $^9_\Lambda$Be($0^+$) and $^9_\Lambda$Be($2^+$) are well reproduced, compared to the experimental value.
However, the use of these interactions leads to the overbinding of the ground state of $^{13}_\Lambda{\rm C}$ by about $3$ MeV, while the original parameter set of odd state part of the $\Lambda$N interaction, shown in the parenthesis in Table.\ref{tab:YNG}, with $k_F=1.062$ fm, gives much better agreement with the experimental data~\cite{Hiyama1997ptep}. This is because the ground state of $^{13}_\Lambda{\rm C}$ has a higher denisty with more compact shell-model-like struture than $^9_\Lambda{\rm Be}$ and the $0_2^+$ state of $^{13}_\Lambda{\rm C}$. We then replace the ground state of $^{13}_\Lambda{\rm C}$ with the one obtained by the more appropriate interaction (see Fig.~\ref{fig:0pspec}).


\section{Results and Discussions}
\subsection{Energy spectra of the $0^+$ states of $^{13}_{\Lambda}{\rm C}$}\label{sec:IIIa}

\begin{figure*}[tb]
\setlength{\abovecaptionskip}{0.cm}
\setlength{\belowcaptionskip}{-0.cm}
\centerline{\includegraphics[width=16.0 cm,height=10.0 cm]
                                              {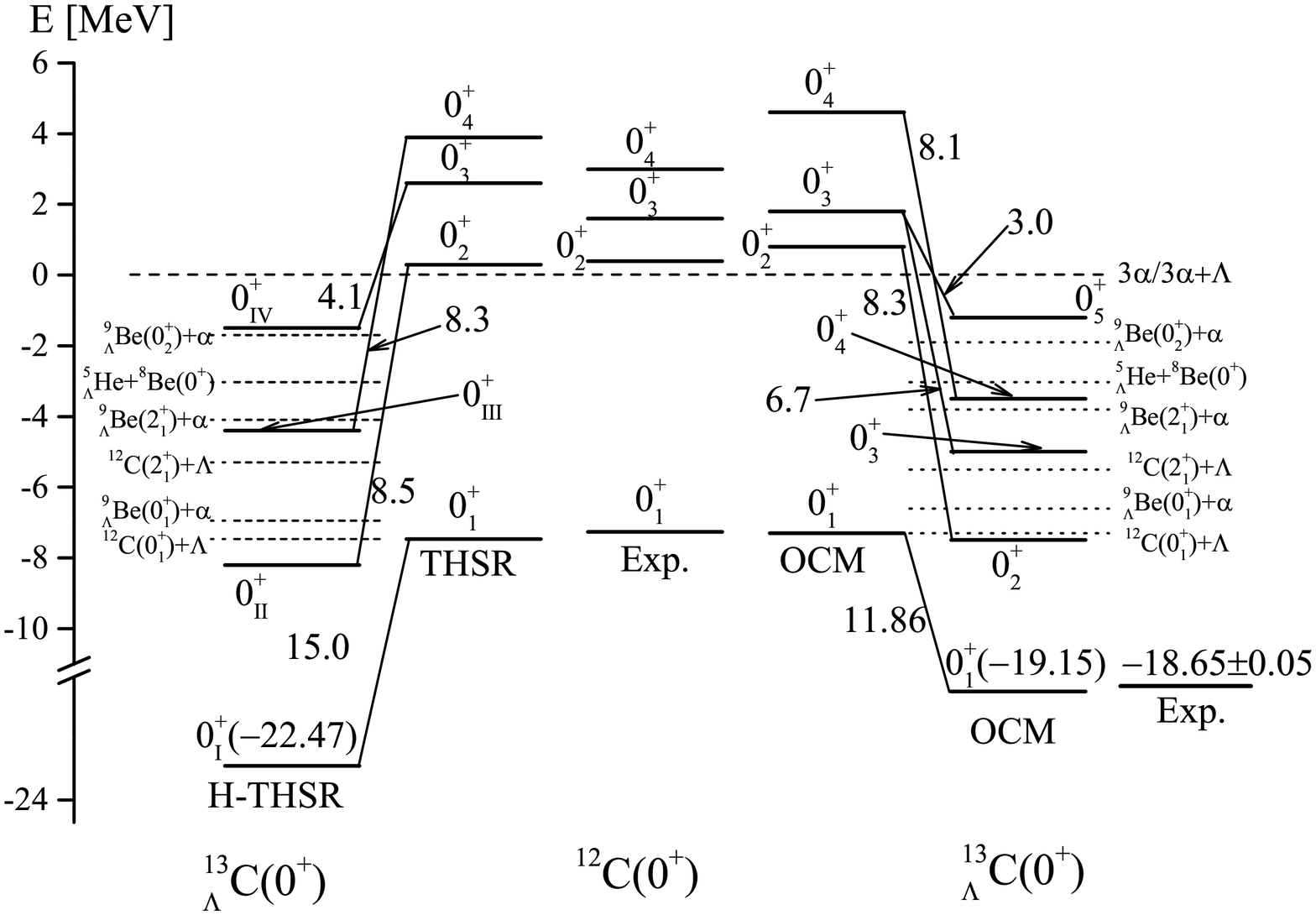}}
\caption{
The calculated energy levels of $^{12}$C and $^{13}_{\Lambda}$C. The left two columns show the $J^\pi=0^+$ spectra of $^{13}_{\Lambda}$C and $^{12}$C, which are calculated with the $3\alpha$ and $3\alpha+\Lambda$ THSR ansatz, respectively~\cite{Funaki2018plb}. The right two columns show the $J^\pi=0^+$ spectra of $^{13}_{\Lambda}$C and $^{12}$C, which are calculated with the $3\alpha$ and $3\alpha+\Lambda$ OCM, respectively. The experimental values of $^{12}$C are shown in the the middle column~\cite{Cook1957,Itoh_2013}.
}
\label{fig:0pspec}
\end{figure*}

In Fig.~\ref{fig:0pspec}, we show the calculated $0^+$ states of $^{13}_{\Lambda}{\rm C}$, together with the $0^+$ states of $^{12}{\rm C}$, which are obtained with the $3\alpha$ OCM with the same $\alpha$-$\alpha$ interaction as used in $^{13}_{\Lambda}{\rm C}$. The results of Ref.~\cite{Funaki2018plb} obtained by using the H-THSR wave function and those of $^{12}{\rm C}$ by the same THSR ansatz in Ref.~\cite{funaki2015hoyleband}, together with the corresponding experimental data, are also shown.

First, we discuss the $0^+$ states of $^{12}{\rm C}$, which are shown in the middle of Fig.~\ref{fig:0pspec}. While the ground state of $^{12}{\rm C}$ has a shell-model-like structure, the $0^+_2$ state, i.e. the Hoyle state, has a gaslike structure of the $3\alpha$ clusters~\cite{HoriuchiOCM,THSR2001,Funaki2003}. The $0^+_3$ state is considered to have a higher nodal structure for the $^8{\rm Be}+\alpha$ relative wave function. It is also emphasized that this state has an analogous structure to the Hoyle state, as excited from the Hoyle state with a large monopole transition strength. On the other hand, the $0_4^+$ state is considered to have a linear-chain-like structure of the $3\alpha$ clusters~Refs.\cite{AMD2007,FMD2007,funaki2016}, which is the different cluster structure from those of the Hoyle and $0^+_3$ states. We consider that all these four $0^+$ states are well reproduced by the present $3\alpha$ OCM calculation. The $0_2^+$, $0_3^+$, and $0_4^+$ states are obtained at $0.8$ MeV, $1.8$ MeV, and $4.9$ MeV above the $3\alpha$ threshold, respectively, within the bound state approximation. The calculated root mean square (rms) radii of these states are shown in Table.~\ref{tab:rms}. The rms radius of the Hoyle state, $4.2$ fm, is much larger than that of the more compact ground state, $2.4$ fm, and is smaller than those of the $0_3^+$ and $0_4^+$ states, $5.6$ fm and $5.4$ fm, respectively. This tendency is consistent with the previous calculations with the THSR ansatz~\cite{funaki2016}.


Let us then discuss the $0^+$ states of $^{13}_{\Lambda}$C, which are also shown in Fig.~\ref{fig:0pspec}. As we mentioned in the previous section, the ground state is calculated with the original parameter sets of $\Lambda N$ interaction~\cite{Yamamoto1994yng}. The calculated binding energy, $E=-19.15$ MeV, measured from the $3\alpha + \Lambda$ threshold, and also $B_\Lambda=11.86$ MeV~\cite{Tamura2006}, are in good agreement with the corresponding experimental values, $E=-18.65\pm0.05$ MeV and $B_\Lambda=11.38\pm0.05$ MeV, respectively. We can also see that the $0_2^+$ state is slightly bound below the lowest $^{12}{\rm C}+\Lambda$ threshold and the other higher three $0^+$ states are obtained as resonant states. The binding energies for the $0_2^+$, $0_3^+$, $0_4^+$, and $0_5^+$ states are calculated as $E=-7.49$ MeV, $E=-5.0$ MeV, $E=-3.5$ MeV, and $E=-1.2$ MeV, respectively. We should note that, as mentioned in Introduction, apart from $^{12}{\rm C}$, many thresholds such as $^{9}_{\Lambda}{\rm Be}(0^+_1)+\alpha$, $^{12}{\rm C}(2^+_1)+\Lambda$, $^{9}_\Lambda{\rm Be}(2^+_1)+\alpha$, $^{5}_{\Lambda}{\rm He}+^8{\rm Be}(0^+)$, and $^{5}_{\Lambda}{\rm He}+2\alpha$, are newly open in $^{13}_{\Lambda}{\rm C}$, which may give a variety of cluster structures.


For comparison, in the left most column of Fig.~\ref{fig:0pspec}, we show the energy spectra of $0^+$ states of $^{13}_{\Lambda}{\rm C}$ calculated with the H-THSR wave function~\cite{Funaki2018plb}, where the four $0^+$ states are obtained below the $3\alpha+\Lambda$ threshold, while we obtain the five $0^+$ states in the present calculation. The authors in Ref.~\cite{Funaki2018plb} show that the $0^+_{\rm {\uppercase\expandafter{\romannumeral1}}}$, $0^+_{\rm {\uppercase\expandafter{\romannumeral2}}}$, $0^+_{\rm {\uppercase\expandafter{\romannumeral3}}}$, and $0^+_{\rm {\uppercase\expandafter{\romannumeral4}}}$ states keep the structures of the $0_1^+$, $0_2^+$, $0_4^+$, and $0_3^+$ states of $^{12}{\rm C}$, to which the $\Lambda$ particle couples, respectively.
In particular, it is argued that the $0^+_{\rm {\uppercase\expandafter{\romannumeral4}}}$ state includes the $^9_\Lambda{\rm Be}(0_2^+)+\alpha$ component dominantly, where the $0_2^+$ state of $^9_\Lambda{\rm Be}$ is identified as a $^5_\Lambda{\rm He} +\alpha$ resonant state. It is thus concluded that the $0^+_{\rm {\uppercase\expandafter{\romannumeral4}}}$ state has an analogous structure to the Hoyle analog state, a gaslike $^5_\Lambda{\rm He} +2\alpha$ cluster structure. However, as we will discuss later, it should be noted that in our calculation with the OCM, we cannot find the $0_2^+$ state of $^9_\Lambda{\rm Be}$ around this energy region~\cite{qian2020}.

\begin{table}[htbp]
\caption{Rms radii of the $0^+$ states of $^{13}_\Lambda$C and $^{12}$C. The units are in fm.}
\begin{tabular}{p{2cm}<{\centering}p{1.2cm}<{\centering}p{1.2cm}<{\centering}p{2cm}<{\centering}
               p{1.2cm}<{\centering}}
  \hline
  \hline
 $^{13}_\Lambda$C & ${\rm R}_{\rm rms}^{{\rm c}-\Lambda}$ & ${\rm R}_{\rm rms}$ & $^{12}$C&  ${\rm R}_{\rm rms}$\\
  \hline
 $^{13}_\Lambda$C$(0_1^+)$ & 2.3   & 2.3   &$^{12}$C$(0_1^+)$  & 2.4 \\
 \hline
 $^{13}_\Lambda$C$(0_2^+)$ & 3.1   & 3.1   &$^{12}$C$(0_2^+)$  & 4.2 \\
 \hline
 $^{13}_\Lambda$C$(0_3^+)$ & 4.3   & 5.4   &$^{12}$C$(0_3^+)$  & 5.6 \\
 \hline
 $^{13}_\Lambda$C$(0_4^+)$ & 4.3   & 4.1   &$^{12}$C$(0_4^+)$  & 5.4 \\
 \hline
 $^{13}_\Lambda$C$(0_5^+)$ & 5.2   & 5.6   & \\
 \hline
\end{tabular}
\label{tab:rms}
\end{table}

\subsection{$0^+_1$, $0^+_2$ and $0^+_4$ in $^{13}_{\Lambda}{\rm C}$}\label{sec:IIIb}
In order to investigate the structures of the obtained $0^+$ states and the analogous nature between $^{13}_{\Lambda}$C and $^{12}$C, we calculate the $S^2$-factors of the states, which are obtained by integrating the square of the RWAs, as follows:
\begin{equation}
S^2_i=\int \mathscr{Y}_i(r)^{2}r^{2}dr,
\end{equation}
where $\mathscr{Y}_i(r)$ are the RWAs defined below,
{\small
\begin{eqnarray}
{\cal Y}_i(r)=
\left\{
\begin{aligned}
&\sqrt{\frac{3!}{3!1!}}\left\langle  \left[
\frac{\delta(\rho_{3}-r)}{\rho_{3}^2}Y_{0}(\hat{\rho}_{3}), \phi_i(^{12}{\rm{C}})
\right]_{00} \bigg{|}\Psi(^{13}_{\Lambda}\rm{C}) \right\rangle,\\ \nonumber
&\qquad\qquad\qquad\qquad\qquad\qquad\qquad\qquad(i=1,\cdots,5),\\ \nonumber
&\sqrt{\frac{3!}{2!1!}}\left\langle \left[
\frac{\delta(\rho_{1}-r)}{\rho_{1}^2}Y_{0}(\hat{\rho}_{1}), \phi_i(^{9}_{\Lambda}{\rm{Be}})
\right]_{00}  \bigg{|} \Psi(^{13}_{\Lambda}\rm{C}) \right\rangle,\\ \nonumber
&\qquad\qquad\qquad\qquad\qquad\qquad\qquad\qquad(i=6, 7, 10), \\ \nonumber
&\sqrt{\frac{3!}{2!1!}}\!\left\langle \left[
\frac{\delta(\rho_4-r)}{\rho_4^{2}}Y_{0}(\hat{\rho}_4),\!\phi(^{5}_{\Lambda}{\rm{He}})
\phi_i(^{8}{\rm{Be}})\right]_{00} \bigg{|} \Psi(^{13}_{\Lambda}\rm{C})\!\right\rangle,\\
&\qquad\qquad\qquad\qquad\qquad\qquad\qquad\qquad(i=8, 9). \label{eq:rwa}
\end{aligned}
\right.
& \\
\end{eqnarray}
}
Here, $\rho_3$, $\rho_1$ and $\rho_4$ are the relative coordinates shown in Fig. \ref{fig:c13jacobi}. $\phi_i(^{12}{\rm{C}})$, with the channel numbers $i=1,\cdots,5$, denote the wave functions of the $0_1^+$, $0^+_2$, $0^+_3$, $0^+_4$ and $2^+_1$ states of $^{12}$C, respectively, which are calculated with the $3\alpha$ OCM. $\phi_i(^{9}_{\Lambda}{\rm{Be}})$ with $i=6, 7, 10$ denote the wave functions of the $0_1^+$, $2^+_1$, and $0_2^+$ states of $^{9}_{\Lambda}{\rm{Be}}$, respectively, which are calculated with the same framework of the $2\alpha+\Lambda$ OCM as the present $3\alpha + \Lambda$ OCM. We will explain the detail of the $0_2^+$ state of ${^9_\Lambda{\rm Be}}$ later in Sec.~\ref{sec:IIIc}. $\phi_i(^{8}{\rm{Be}})$ with $i=8, 9$ denote the wave functions of the $0^+$ and $2^+$ states of $^8$Be, respectively, which are calculated with the $2\alpha$ OCM. The $\Psi(^{13}_{\Lambda}\rm{C})$ is the wave function of the $0^+$ state of $^{13}_\Lambda {\rm C}$ in Eq.~(\ref{eq:totalwf}).

\begin{figure}[htbp]
\setlength{\abovecaptionskip}{0.cm}
\setlength{\belowcaptionskip}{-0.cm}
\centering
\includegraphics[width=0.46\textwidth]{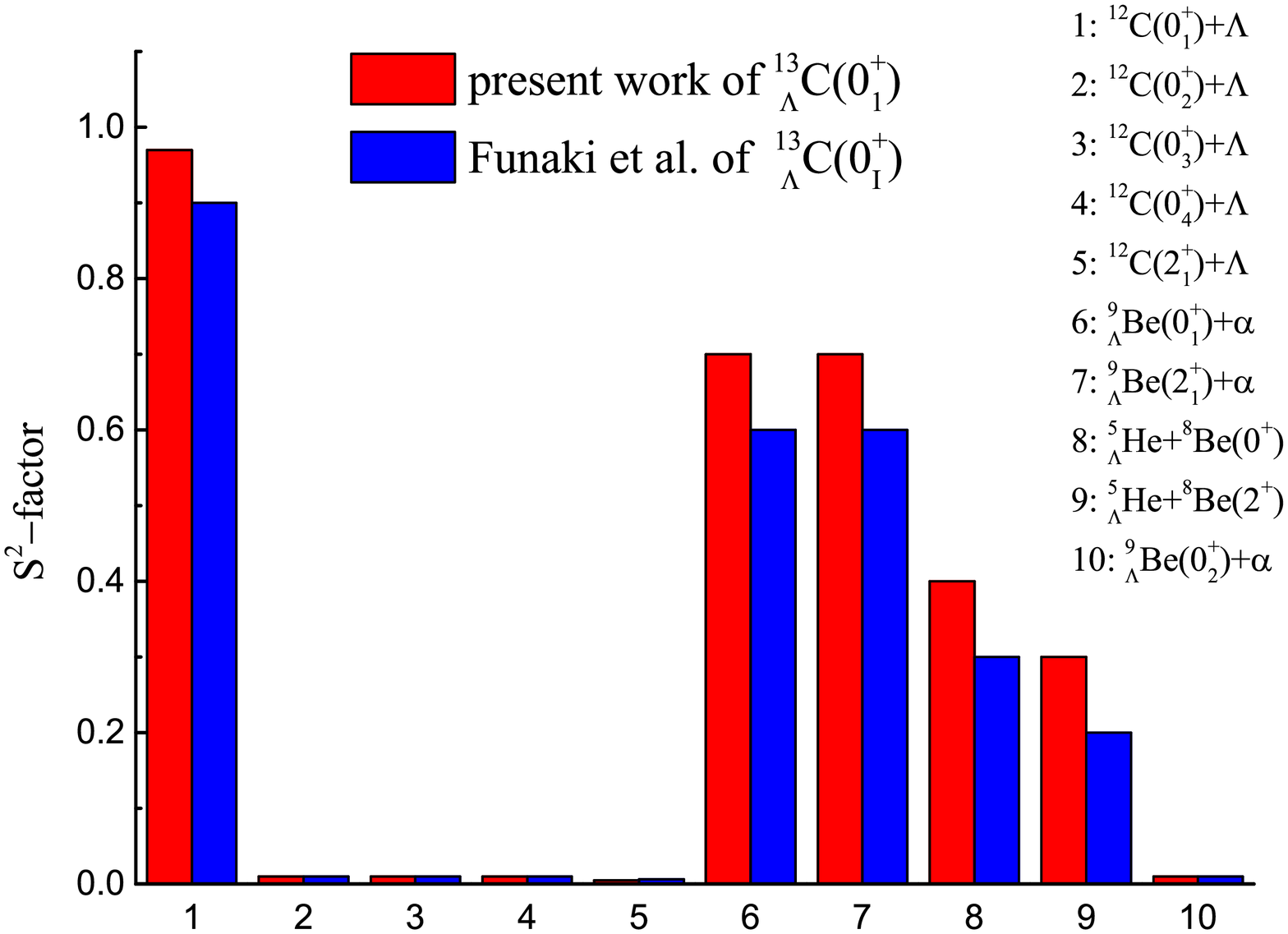}
\caption{
The $S^2$-factors of the $0_1^+$ state in the present calculation and of the $0^+_{\rm {\uppercase\expandafter{\romannumeral1}}}$ state in Fig.~\ref{fig:0pspec} calculated with the H-THSR ansatz~\cite{Funaki2018plb}, are shown by red and blue bars, respectively. The labels 1,2,3,4, and 5 represent the $^{12}{\rm C}(0^+_1)+\Lambda$, $^{12}{\rm C}(0^+_2)+\Lambda$, $^{12}{\rm C}(0^+_3)+\Lambda$, $^{12}{\rm C}(0^+_4)+\Lambda$, and $^{12}{\rm C}(2^+_1)+\Lambda$ channels, respectively, and the labels 6, 7, 8, 9, 10 represent the $^{9}_\Lambda{\rm Be}(0^+_1)+\alpha$, $^{9}_\Lambda{\rm Be}(2^+_1)+\alpha$, $^{5}_\Lambda{\rm He}+^8{\rm Be}(0^+)$, $^{5}_\Lambda{\rm He}+{^8{\rm Be}(2^+)}$, $^{9}_\Lambda{\rm Be}(0^+_2)+\alpha$ channels, respectively. }
\label{fig:sf-1}
\end{figure}

\begin{figure}[htbp]
\setlength{\abovecaptionskip}{0.cm}
\setlength{\belowcaptionskip}{-0.cm}
\centering
\includegraphics[width=0.46\textwidth]{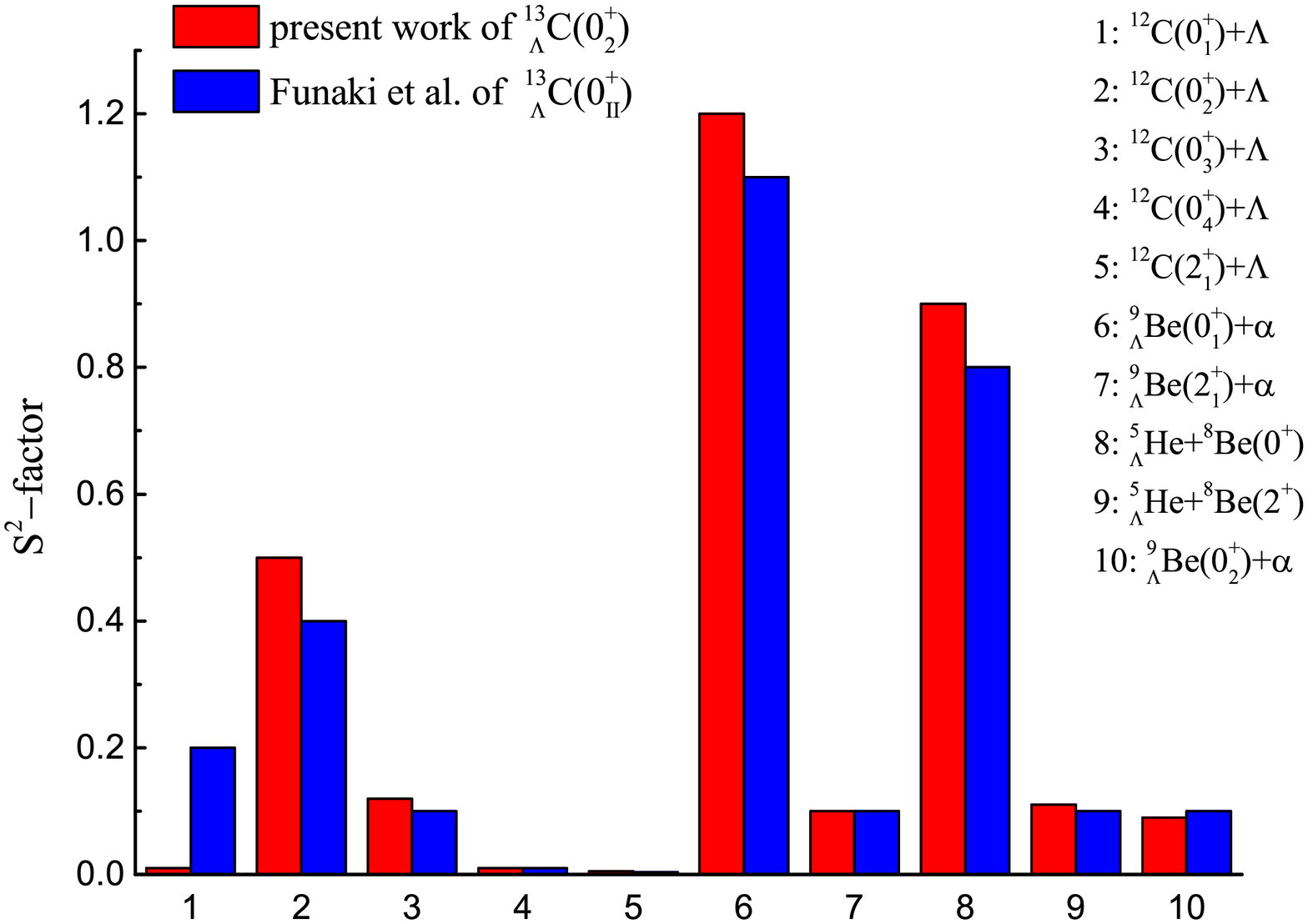}
\caption{
The $S^2$-factors of the $0_2^+$ state in the present calculation and of the $0^+_{\rm {\uppercase\expandafter{\romannumeral2}}}$ state in Fig.~\ref{fig:0pspec} calculated with the H-THSR ansatz~\cite{Funaki2018plb}, are shown by red and blue bars, respectively. The labels 1,2,3,4, and 5 represent the $^{12}{\rm C}(0^+_1)+\Lambda$, $^{12}{\rm C}(0^+_2)+\Lambda$, $^{12}{\rm C}(0^+_3)+\Lambda$, $^{12}{\rm C}(0^+_4)+\Lambda$, and $^{12}{\rm C}(2^+_1)+\Lambda$ channels, respectively, and the labels 6, 7, 8, 9, 10 represent the $^{9}_\Lambda{\rm Be}(0^+_1)+\alpha$, $^{9}_\Lambda{\rm Be}(2^+_1)+\alpha$, $^{5}_\Lambda{\rm He}+^8{\rm Be}(0^+)$, $^{5}_\Lambda{\rm He}+{^8{\rm Be}(2^+)}$, $^{9}_\Lambda{\rm Be}(0^+_2)+\alpha$ channels, respectively.}
\label{fig:sf-2}
\end{figure}

\begin{figure}[htbp]
\setlength{\abovecaptionskip}{0.cm}
\setlength{\belowcaptionskip}{-0.cm}
\centering
\includegraphics[width=0.46\textwidth]{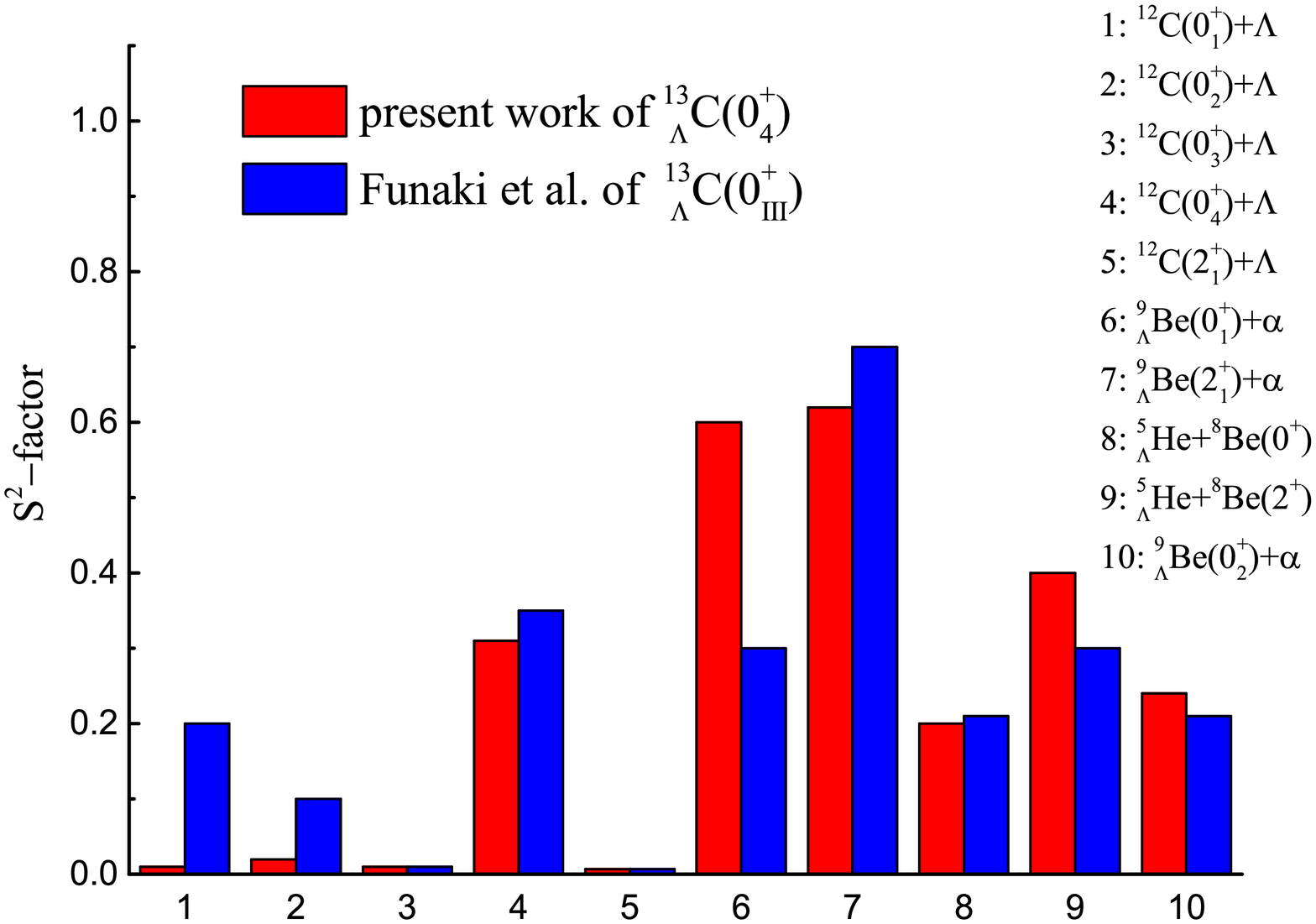}
\caption{
The $S^2$-factors of the $0_4^+$ state in the present calculation and of the corresponding $0^+_{\rm {\uppercase\expandafter{\romannumeral3}}}$ state in Fig.~\ref{fig:0pspec} calculated with the H-THSR ansatz~\cite{Funaki2018plb}, are shown by red and blue bars, respectively. The labels 1,2,3,4, and 5 represent the $^{12}{\rm C}(0^+_1)+\Lambda$, $^{12}{\rm C}(0^+_2)+\Lambda$, $^{12}{\rm C}(0^+_3)+\Lambda$, $^{12}{\rm C}(0^+_4)+\Lambda$, and $^{12}{\rm C}(2^+_1)+\Lambda$ channels, respectively, and the labels 6, 7, 8, 9, 10 represent the $^{9}_\Lambda{\rm Be}(0^+_1)+\alpha$, $^{9}_\Lambda{\rm Be}(2^+_1)+\alpha$, $^{5}_\Lambda{\rm He}+{^8{\rm Be}(0^+)}$, $^{5}_\Lambda{\rm He}+{^8{\rm Be}(2^+)}$, $^{9}_\Lambda{\rm Be}(0^+_2)+\alpha$ channels, respectively.}
\label{fig:sf-4}
\end{figure}

\begin{figure}[htbp]
\setlength{\abovecaptionskip}{0.cm}
\setlength{\belowcaptionskip}{-0.cm}
\centering
\includegraphics[width=0.46\textwidth]{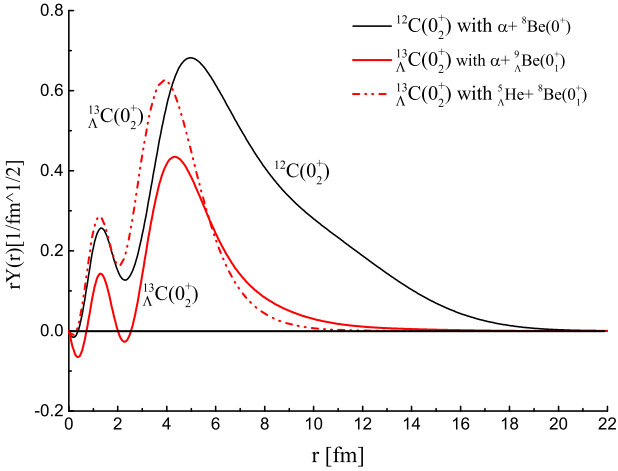}
\caption{
The RWAs of the $0_2^+$ state of $^{12}{\rm C}$ in the $^{8}{\rm Be}(0^+)+\alpha$ channel is shown in the black solid line. The red solid and red dash-dotted lines represent the RWAs of the $0_2^+$ state of $^{13}_{\Lambda}{\rm C}(0^+_2)$ in the $^{9}_\Lambda{\rm Be}(0^+_1)+\alpha$ and $^{5}_\Lambda{\rm He}+^8{\rm Be}(0^+)$ channels, respectively.}
\label{fig:2nd0prwa}
\end{figure}

The calculated $S^2$-factors of the $0_1^+$, $0_2^+$, and $0_4^+$ states are shown in Fig.~\ref{fig:sf-1}, Fig.~\ref{fig:sf-2} and Fig.~\ref{fig:sf-4}, respectively, colored in red, for the 10 channels. For comparison, we also show the $S^2$-factors calculated by Funaki et al.~\cite{Funaki2018plb} side by side, colored in blue.


We can see that in Fig.~\ref{fig:sf-1}, the $0_1^+$ state has the largest fraction from the $^{12}$C$(0^+_1)+\Lambda$ channel (channel 1), which means that the $0^+_1$ state dominantly has the structure that the $\Lambda$ particle couples to the $^{12}$C$(0^+_1)$ core in an $S$-wave, keeping the shell-model-like structure. This state also has large components from the channels $^{9}_{\Lambda}{\rm Be}(0^+_1)+\alpha$ (channel 6) and $^{9}_{\Lambda}{\rm Be}(2^+_1)+\alpha$ (channel 7), which again show that it keeps the SU(3)-like nature of the $0_1^+$ state of $^{12}{\rm C}$, with the $\Lambda$ particle coupling to $^8{\rm Be}$ core in an $S$-wave. All these distributions are consistent with those obtained in Ref.~\cite{Funaki2018plb} colored in blue. We also mention that the rms radius, which is shown in Table~\ref{tab:rms} for this state, $2.3$ fm is almost the same as that for the $0_1^+$ state of $^{12}{\rm C}$, $2.4$ fm. This indicates that there is almost no shrinkage by adding the $\Lambda$ particle to the $^{12}$C$(0^+_1)$ core, since the density of the ground state of $^{12}{\rm C}$ is close to saturation.

The dominant channels for the $0^+_2$ state of $^{13}_{\Lambda}$C are $^{9}_{\Lambda}{\rm Be}(0^+_1)+\alpha$ (channel 6) and $^{5}_{\Lambda}{\rm He}+^8{\rm Be}(0^+)$ (channel 8), and the channel $^{12}{\rm C}(0^+_2)+\Lambda$ (channel 2) gives smaller contribution, although the component of this channel is the largest among the five $^{12}{\rm C}+\alpha$ channels (channels 1,2,3,4,5). On the other hand, we can see that in Table~\ref{tab:rms}, the rms radius of this state, $3.1$ fm, is much smaller than that of the Hoyle state, $4.2$ fm. This is because the density of the Hoyle state is much lower than saturation, and therefore the $\Lambda$ particle can shrink the Hoyle state so much to be the more compact $0_2^+$ state of $^{13}_\Lambda{\rm C}$. This core shrinkage gives smaller overlap between the $0_2^+$ state of $^{13}_\Lambda{\rm C}$ and $^{12}{\rm C}(0^+_2)+\Lambda$ in the correponding RWA, leading to the smaller $S^2$-factor in this channel. It is also interesting to consider the energy gain between the $0_2^+$ state of $^{13}_{\Lambda}$C and the Hoyle state, $B_\Lambda=8.3$ MeV, which is $2.7$ times larger than $B_\Lambda=3.12$ MeV for $^5_\Lambda{\rm He}$ and is larger than $B_\Lambda=6.71$ MeV for $^9_\Lambda{\rm Be}$. This indicates that the $\Lambda$ particle strongly couples to the $^8{\rm Be}$ core as well as the $^4{\rm He}$ core, i.e. all the $\alpha$ clusters, which is consistent with the fact that this state has large $S^2$-factors from the channels 6 and 8. We should also mention that the distributions of the $S^2$-factors in the present calculations and in Ref.~\cite{Funaki2018plb} are in good agreement with each other.

In Fig.~\ref{fig:2nd0prwa}, we show the RWAs of the $0_2^+$ state of $^{13}_{\Lambda}{\rm C}$ in the $^{9}_\Lambda{\rm Be}(0^+_1)+\alpha$ (${\cal Y}_{i=6}(r)$) and $^{5}_\Lambda{\rm He}+{^8{\rm Be}(0^+)}$ (${\cal Y}_{i=8}(r)$) channels. For comparison, we also show the RWAs of the Hoyle state in the $^{8}{\rm Be}(0^+)+\alpha$ channel. The RWAs for $^{13}_{\Lambda}{\rm C}$ in both channels are pulled into inner region, in comparison with that for the Hoyle state, due to the shrinkage effect of the $\Lambda$ particle. However, the behaviours of oscillation in inner region of the RWA in the $^{5}_\Lambda{\rm He}+^8{\rm Be}(0^+)$ channel and the one for the Hoyle state in the $^{8}{\rm Be}(0^+)+\alpha$ channel are similar to each other. In both cases, the nodes in the inner region disappear and change to the oscillation in a similar way, since the core $^8{\rm Be}$ dissolves into $2\alpha$ clusters and the $\alpha$ or $^5_\Lambda{\rm He}$ cluster does not feel the effect of the antisymmetrization from the core sufficiently. On the other hand, the three nodes recover for the RWA in the $^{9}_\Lambda{\rm Be}(0^+_1)+\alpha$ channel, since the core $^{9}_\Lambda{\rm Be}(0^+_1)$ is a compact object, again due to the shrinkage effect of the $\Lambda$ particle.



Let us then discuss the $0^+_4$ state. The $S^2$-factors for this state are shown in Fig.~\ref{fig:sf-4}, denoted by red bars. In comparison with them, we also show those for the state calculated as the $0^+_{\rm {\uppercase\expandafter{\romannumeral3}}}$ state with the H-THSR ansatz by blue bars. Although the component of $^9_\Lambda{\rm Be}(0_1^+)+\alpha$ channel (channel 6) in our calculation is about 2 times larger than in the H-THSR calculation, both distributions are in good agreement with each other, and hence we can say that our $0_4^+$ state corresponds to the $0^+_{\rm {\uppercase\expandafter{\romannumeral3}}}$ state in the H-THSR calculation, which is shown to have a structure that the $\Lambda$ particle moves around a linear-chain-like configuration of the $3\alpha$ clusters in Ref.~\cite{Funaki2018plb}. The largest component of the $^{12}{\rm C}+\Lambda$ channels is of $^{12}{\rm C}(0^+_4)+\alpha$ channel. Many previous cluster model calculations show that the $0_4^+$ state of $^{12}{\rm C}$ dominantly has the $^8{\rm Be}(2_1^+)+\alpha$ component and this state is considered to have a linear-chain-like structure, as mentioned in Introduction~\cite{Fujiwara1980-hoyleband,AMD2007,FMD2007}. We can see that the present $0_4^+$ state of $^{13}_\Lambda{\rm C}$ also has the largest component from the $^9_\Lambda{\rm Be}(2_1^+)+\alpha$ channel as well as large component from the $^{5}_{\Lambda}{\rm He}+^8{\rm Be}(2^+)$ channel, which also supports the idea that in our $0_4^+$ state the $\Lambda$ particle moves in an $S$-wave around the $0_4^+$ state of $^{12}{\rm C}$ with the linear-chain-like configuration.



Concerning the shrinkage effect between the core $^{12}$C and the $^{13}_{\Lambda}$C, as shown in Table.~\ref{tab:rms}, the density of the $0_4^+$ state is $(5.4{\rm fm}/4.1{\rm fm})^3=2.3$ times smaller than the $0_4^+$ state of $^{12}{\rm C}$. This amount of shrinkage is similar to that between the $0_2^+$ state and the Hoyle state in $^{12}{\rm C}$, $(4.2{\rm fm}/3.1{\rm fm})^3=2.5$. The energy gain of this state from the $0_4^+$ state of $^{12}{\rm C}$, $8.1$ MeV, is also close to that of the $0_2^+$ state from the Hoyle state, which is $8.3$ MeV. Thus as in the same way as the $0_2^+$ state, in the $0_4^+$ state, the $\Lambda$ particle couples to all of the $\alpha$ clusters, keeping a linear-chain-like arrangement of the $3\alpha$ clusters.


\subsection{$0^+_3$ and $0^+_5$ in $^{13}_{\Lambda}{\rm C}$}\label{sec:IIIc}

\begin{figure}[htbp]
\setlength{\abovecaptionskip}{0.cm}
\setlength{\belowcaptionskip}{-0.cm}
\centering
\includegraphics[width=0.46\textwidth]{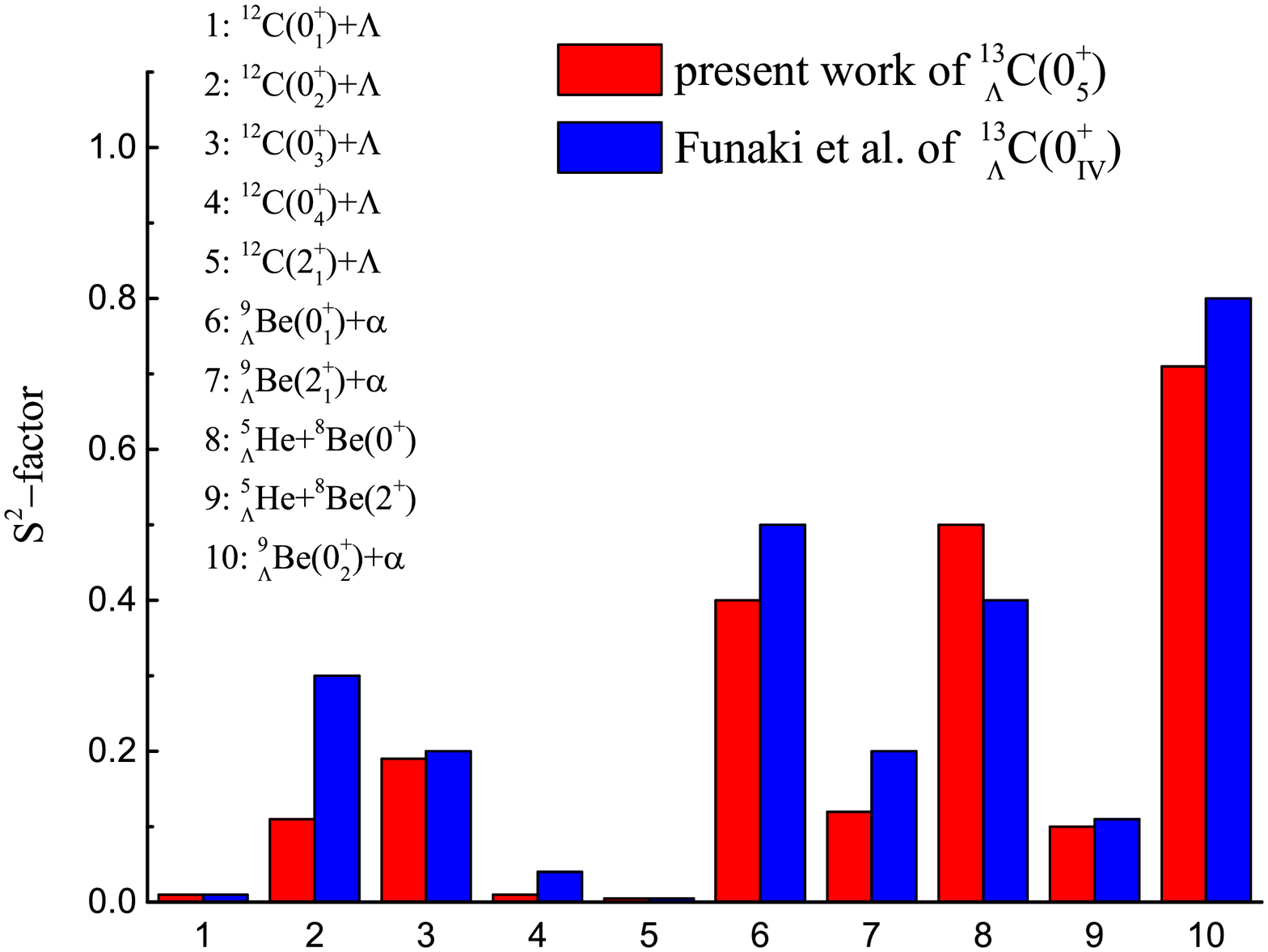}
\caption{
The $S^2$-factors of the $0_5^+$ state in the present calculation and of the $0^+_{\rm {\uppercase\expandafter{\romannumeral4}}}$ state in Fig.~\ref{fig:0pspec} calculated with the H-THSR ansatz~\cite{Funaki2018plb}, are shown by red and blue bars, respectively. The labels 1,2,3,4, and 5 represent the $^{12}{\rm C}(0^+_1)+\Lambda$, $^{12}{\rm C}(0^+_2)+\Lambda$, $^{12}{\rm C}(0^+_3)+\Lambda$, $^{12}{\rm C}(0^+_4)+\Lambda$, and $^{12}{\rm C}(2^+_1)+\Lambda$ channels, respectively, and the labels 6, 7, 8, 9, 10 represent the $^{9}_\Lambda{\rm Be}(0^+_1)+\alpha$, $^{9}_\Lambda{\rm Be}(2^+_1)+\alpha$, $^{5}_\Lambda{\rm He}+^8{\rm Be}(0^+)$, $^{5}_\Lambda{\rm He}+{^8{\rm Be}(2^+)}$, $^{9}_\Lambda{\rm Be}(0^+_2)+\alpha$ channels, respectively.}
\label{fig:sf-5}
\end{figure}

In Fig.~\ref{fig:sf-5}, we show the $S^2$-factors for the $0_5^+$ state, which is denoted by red bars, and for the $0^+_{\rm {\uppercase\expandafter{\romannumeral4}}}$ state obtained in Ref.~\cite{Funaki2018plb} with the H-THSR ansatz, for comparison, by blue bars, side by side. We can see that both the distributions are qualitatively in good agreement with each other, and therefore we can conclude that our $0_5^+$ state corresponds to the $0_{\rm IV}^+$ state obtained in Ref.~\cite{Funaki2018plb}. As mentioned in Introduction, the $0_{\rm IV}^+$ state in Ref.~\cite{Funaki2018plb} can be considered to be the Hoyle analog state, where the $\Lambda$ particle strongly couples with an $\alpha$ cluster to form $^5_\Lambda{\rm He}$ and mutually $2\alpha$ and $^5_\Lambda{\rm He}$ clusters loosely couple like a gas. The authors in Ref.~\cite{Funaki2018plb} also discussed the existence of the $0_2^+$ state of $^9_\Lambda{\rm Be}$, which is a two-body resonant state of a well developed $^5_\Lambda{\rm He}$ and $\alpha$ cluster structure, though experimentally the state is still unknown. Thus, the fact that the $0_{\rm IV}^+$ state of $^{13}_\Lambda{\rm C}$ in Ref.~\cite{Funaki2018plb} dominantly has the $^{9}_{\Lambda}{\rm Be}(0^+_2)+\alpha$ channel (channel 10) component, as shown in Fig.~\ref{fig:sf-5}, as well as a large component from the ${^5_\Lambda{\rm He}}+{^8{\rm Be}(0^+)}$ channel (channel 8), gives the evidence that this state has a gaslike $^5_\Lambda{\rm He}+2\alpha$ cluster structure as the Hoyle analog state.

\begin{figure}[htbp]
\setlength{\abovecaptionskip}{0.cm}
\setlength{\belowcaptionskip}{-0.cm}
\centering
\includegraphics[width=0.46\textwidth]{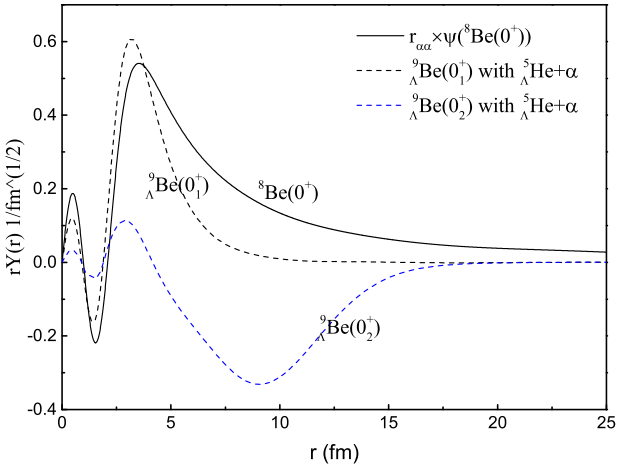}
\caption{
The RWAs of the $0_1^+$ and $0_2^+$ state of $^{9}_\Lambda{\rm Be}$ in the $^{5}_\Lambda{\rm He}+\alpha$ channel are shown in black and blue dashed lines, respectively. The solid line represents the $\alpha$-$\alpha$ relative wave function of the $0^+$ state of $^8{\rm Be}$. See the text for the details of the $0_2^+$ state of $^9_\Lambda{\rm Be}$.}
\label{fig:be92nd0p}
\end{figure}

However, in the present full-channel three-body $\alpha$-$\alpha$-$\Lambda$ calculation by the OCM, we could not obtain the $0_2^+$ state of $^9_\Lambda{\rm Be}$ around the same energy region as they obtained. Nevertheless, we could obtain a $0_2^+$ state of $^{9}_{\Lambda}{\rm Be}$ at $0.7$ MeV below the $2\alpha+\Lambda$ threshold, where the limited model space including $^5_\Lambda{\rm He}+\alpha$ channel are only taken into account~\cite{qian2020}.
In Fig.~\ref{fig:be92nd0p}, we show the RWA of this artificial $0_2^+$ state of $^9_\Lambda{\rm Be}$ in the $^{5}_\Lambda{\rm He}+\alpha$ channel (blue dotted line). For comparison, we also show the RWA of the $0_1^+$ state of $^9_\Lambda{\rm Be}$ in the same channel (black dotted line) and the $\alpha$-$\alpha$ relative wave function of the $0^+$ state of $^8{\rm Be}$ (black solid line). The largest amplitude for the $^9_\Lambda{\rm Be}(0_2^+)$ spreads outside up to around $10$ fm, which indicates that this state has a very well developed $^5_\Lambda{\rm He}+\alpha$ cluster structure. We thus make use of this state as the $^9_\Lambda{\rm Be}(0_2^+)+\alpha$ channel (channel 10). Figure~\ref{fig:sf-5} shows that this channel component gives the largest contribution to the $S^2$-factors for the $0_5^+$ state, in the same way as the $0_{\rm IV}^+$ state of Ref.~\cite{Funaki2018plb}, placed side by side, by blue bars. We can also see that the second largest component is from the ${^5_\Lambda{\rm He}} + {^8{\rm Be}(0^+)}$ channel (channel 8). Therefore, we can say again that our $0_5^+$ state corresponds to the $0_4^+$ state in Ref.~\cite{Funaki2018plb} and this state has the structure that the $^5_\Lambda{\rm He}$ and $2\alpha$ clusters weakly interact with each other in relative $S$ waves, like a gas.

 Besides, among the five $^{12}{\rm C}+\Lambda$ channels, this state has the largest component from the $^{12}{\rm C}(0_3^+)+\Lambda$ channel. As far as both the $0_5^+$ state of $^{13}_\Lambda{\rm C}$ and the $0_3^+$ state of $^{12}{\rm C}$ can be associated with each other, the energy gain of the former state from the latter is $3.0$ MeV, as shown in Fig.~\ref{fig:0pspec}, which is close to the binding energy of $^5_\Lambda{\rm He}$, $3.12$ MeV. The calculated rms radius of this state, $5.6$ fm, is as large as the $0_3^+$ state of $^{12}{\rm C}$, as shown in Table~\ref{tab:rms}. The $\Lambda$ particle namely gives no shrinkage of the core ${^{12}{\rm C}}$ nucleus. All these results again indicate that the $\Lambda$ particle couples with only one $\alpha$ cluster to form $^5_\Lambda{\rm He}$, without any shrinkage of the core, and the $^5_\Lambda{\rm He}$ and $2\alpha$ clusters move freely like a gas, analogous to the dilute $3\alpha$ clustering of the Hoyle state. It is also interesting to mention that the $0_3^+$ state of $^{12}{\rm C}$, which is likely to be a precursor of the Hoyle state, is substituted for the Hoyle state, as the $^5_\Lambda{\rm He}+2\alpha$ gas, when the $\Lambda$ particle is added, while the Hoyle state is converted to the $0_2^+$ state of $^{13}_\Lambda{\rm C}$ with much more compact $^9_\Lambda{\rm Be}+\alpha$ structure.

\begin{figure}[htbp]
\setlength{\abovecaptionskip}{0.cm}
\setlength{\belowcaptionskip}{-0.cm}
\centering
\includegraphics[width=0.46\textwidth]{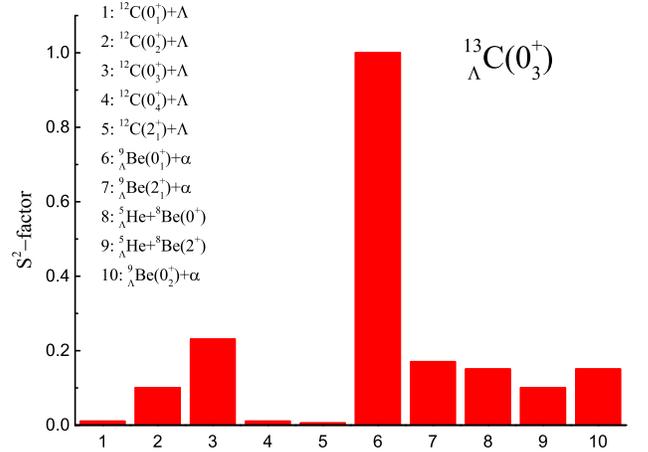}
\caption{
The $S^2$-factors of the $0_3^+$ state in the present calculation are shown. The labels 1,2,3,4, and 5 represent the $^{12}{\rm C}(0^+_1)+\Lambda$, $^{12}{\rm C}(0^+_2)+\Lambda$, $^{12}{\rm C}(0^+_3)+\Lambda$, $^{12}{\rm C}(0^+_4)+\Lambda$, and $^{12}{\rm C}(2^+_1)+\Lambda$ channels, respectively, and the labels 6, 7, 8, 9, 10 represent the $^{9}_\Lambda{\rm Be}(0^+_1)+\alpha$, $^{9}_\Lambda{\rm Be}(2^+_1)+\alpha$, $^{5}_\Lambda{\rm He}+{^8{\rm Be}(0^+)}$, $^{5}_\Lambda{\rm He}+{^8{\rm Be}(2^+)}$, $^{9}_\Lambda{\rm Be}(0^+_2)+\alpha$ channels, respectively.}
\label{fig:sf-3}
\end{figure}

Next let us discuss the $0^+_3$ state. We should mention that this is a new state and has never been obtained in Ref.~\cite{Funaki2018plb} as well as in any other papers. In Fig.\ref{fig:sf-3}, we show the $S^2$-factors for this state. We can see that the component of the $^9_{\Lambda}{\rm Be}(0^+_1)+\alpha$ channel is prominently large. This clearly means that this state dominantly has the $^9_{\Lambda}{\rm Be}(0^+_1)+\alpha$ cluster structure. Very large rms radius of this state, $5.4$ fm, as shown in Table~\ref{tab:rms}, suggests that the $^9_\Lambda{\rm Be}(0_1^+)+\alpha$ cluster structure is very much developed in this state. On the other hand, it should also be mentioned that the $^{12}{\rm C}(0^+_3)+\Lambda$ channel gives the second largest contribution to the $S^2$-factors, indicating that this state can most likely be associated with the $0_3^+$ state of $^{12}{\rm C}$, in a sense that the $\Lambda$ particle can be added to the $^{12}{\rm C}$ core as just an impurity. This assumption of the correspondence with the $0_3^+$ state of $^{12}{\rm C}$ is also consistent with the energy gain from the $0^+_3$ state of $^{12}{\rm C}$, $6.7$ MeV (see Fig.~\ref{fig:0pspec}), which coincides with the $\Lambda$ binding energy for $^9_\Lambda{\rm Be}$, $B_\Lambda=6.71$ MeV. Namely, the $0_3^+$ state of $^{13}_\Lambda{\rm C}$ is given by coupling the $\Lambda$ particle with the $^8{\rm Be}$ part in the $^{12}{\rm C}(0_3^+)$ core, to form the well developed $^9_\Lambda{\rm Be}+\alpha$ cluster structure. Thus we propose the idea that the existence of the $0_3^+$ state of $^{12}{\rm C}$ gives birth to two types of cluster states of $^{13}_\Lambda{\rm C}$, the $0_5^+$ and $0_3^+$ states, where the $\Lambda$ particle couples with only the one $\alpha$ cluster to form $^5_\Lambda{\rm He}$ for the former state and with the two $\alpha$ clusters to form $^9_\Lambda{\rm Be}$ for the latter.

\begin{figure}[htbp]
\setlength{\abovecaptionskip}{0.cm}
\setlength{\belowcaptionskip}{-0.cm}
\centering
\includegraphics[width=0.46\textwidth]{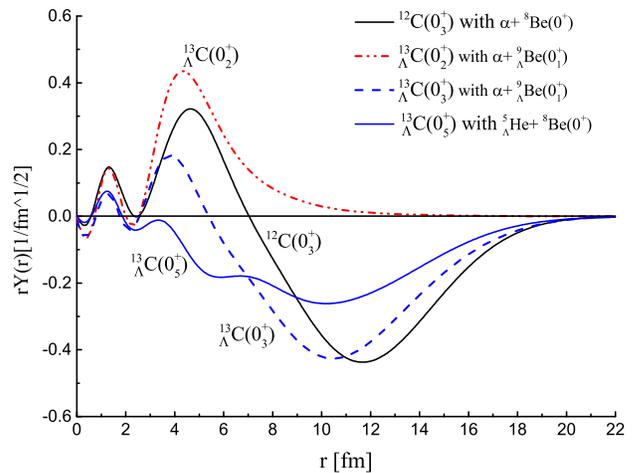}
\caption{
The RWAs of the $0_3^+$ state of $^{12}{\rm C}$ in the $^{8}{\rm Be}(0^+)+\alpha$ channel are shown in black solid line. The red dash-dotted and blue dashed lines represent the RWAs of the $0_2^+$ and $0_3^+$ states of $^{13}_{\Lambda}{\rm C}$ in the $^{9}_\Lambda{\rm Be}+\alpha$ channel, respectively. The blue solid line is the RWAs of the $0_5^+$ state of $^{13}_{\Lambda}{\rm C}$ in the $^8{\rm Be}(0^+)+^{5}_\Lambda{\rm He}$ channel.
}
\label{fig:rwa}
\end{figure}

In order to investigate more deeply all these structures of the $0_3^+$ and $0_5^+$ states, we show in Fig.~\ref{fig:rwa} the RWAs of these states in their important channels, together with relevant RWAs of the $0_2^+$ state of $^{12}{\rm C}$ and the $0_3^+$ state of ${^{13}_\Lambda{\rm C}}$. For the $0^+_3$ state of $^{13}_{\Lambda}$C, the RWA of $^9_\Lambda{\rm Be}+\alpha$ channel, ${\cal Y}_{i=6}(r)$ in Eq.~(\ref{eq:rwa}), is denoted by a blue dashed line. It resembles the RWA in the $^8{\rm Be}+\alpha$ channel of the $0_3^+$ state of $^{12}{\rm C}$, denoted by a black solid line, while the former is drawn into an inner region, compared with the latter, because of a shrinkage effect of the additional $\Lambda$ particle. This clearly indicates that the $0_3^+$ state takes over a higher nodal structure of the relative wave function between the $^8{\rm Be}$ and $\alpha$ cluster for the $0_3^+$ state of $^{12}{\rm C}$. In fact, this RWA has four nodes and hence one more oscillation than the one of the $0_2^+$ state in the $^{9}_\Lambda$Be+$\alpha$ channel (red dash-dotted line). That is why the $0_3^+$ state has the larger rms radius than that of the $0_2^+$ state. These results again support our idea that in the $0_3^+$ state the $\Lambda$ particle is strongly coupled with the $^8{\rm Be}$ part in the $0_3^+$ state of $^{12}{\rm C}$ to form the higher nodal $^9_\Lambda{\rm Be}+\alpha$ cluster structure, due to the orthogonalization to the $0_2^+$ state, with the large component of $^9_\Lambda{\rm Be}+\alpha$ cluster structure.

On the other hand, in the $0_5^+$ state the RWA in the ${^5_\Lambda{\rm He}}+{^8{\rm Be}(0^+)}$ channel, ${\cal Y}_{i=8}(\rho)$ in Eq.~(\ref{eq:rwa}), which is denoted by blue solid line, has a different nodal behavior, in which the nodes disappear in an outer region more than about $2$ fm and only a few small oscillations remain. Since the outer most nodal position corresponds to a range of the core giving the effect of antisymmetrization to the remaining cluster, this result means that the $^5_\Lambda{\rm He}$ cluster does not feel the effect of the antisymmetrization from the $^8{\rm Be}$ core. In other words, the core $^8{\rm Be}$ is dissolved into the $2\alpha$ clusters, and a dilute $^5_\Lambda{\rm He}+2\alpha$ cluster structure is realized.

\section{Summary}\label{sec:IV}

We investigate the structures of the $0^+$ states in $^{13}_{\Lambda}$C by using the $3\alpha+\Lambda$ OCM, which is the semi-microscopic four-body cluster model. In order to solve the Schr\"odinger equation of the $3\alpha+\Lambda$ four-body system, we apply the Gaussian expansion method. We use the $\alpha$-$\alpha$ interaction, which reproduces the observed $\alpha$-$\alpha$ scattering data, and an effective three-body repulsive force, in order to fit the binding energy of the ground state of $^{12}$C to the empirical value. The Pauli forbidden states $(0s,1s,0d)$ between the $2\alpha$ clusters are ruled out by introducing the Pauli exclusion operator in the OCM. We employed the $\alpha \Lambda$ potential by folding procedure of YNG-NF $\Lambda N$ interaction with the $\alpha$ cluster wave function with $(0s)^4$ configuration. We adjust even- and odd-states of the $\Lambda N$ interaction, so as to reproduce the observed binding energies of ground states of $^5_{\Lambda}$He and $^9_{\Lambda}$Be. We then obtain the five $0^+$ states.

In order to investigate the structures of the five $0^+$ states and discuss the binding mechanism between the $\Lambda$ particle and core nucleus, we calculate the energy spectrum, rms radii, the RWAs, and $S^2$-factors of these states. The results are compared with the ones calculated by Funaki et al. in Ref.~\cite{Funaki2018plb} by using the so-called H-THSR ansatz, where the four $0^+$ states are obtained.

The $0_1^+$ state is analogous to the ground state of $^{12}{\rm C}$, where the structure of the $^{12}{\rm C}$ core in $^{13}_{\Lambda}{\rm C}$ is almost the same as the one in the ground state of $^{12}$C. There is no shrinkage effect between the $0_1^+$ state of ${^{13}_{\Lambda}{\rm C}}$ and the ground state of $^{12}$C. The $0_2^+$ state is found to correspond to the Hoyle state, where the $\Lambda$ particle couples strongly with all the $3\alpha$ clusters and causes an outstanding reduction of the spatial size.

The $0_4^+$ state in $^{13}_{\Lambda}$C corresponds to the $0_4^+$ state of $^{12}{\rm C}$, which has a linear-chain-like structure. We find that the $D$-wave component between $^9_\Lambda{\rm Be}$ and $\alpha$ cluster is largely included in this state, which indicates that this state may still keep a linear-chain-like structure as the $0_4^+$ state of $^{12}{\rm C}$.

We find that the $0_5^+$ state may correspond to the $0_3^+$ state of $^{12}{\rm C}$ and has a large $S^2$-factor in the $^{9}_{\Lambda}{\rm Be}(0^+_2)+\alpha$ channel. Considering the fact that the $0_2^+$ state of $^{9}_{\Lambda}{\rm Be}$ has the well developed $^5_\Lambda{\rm He} + \alpha$ cluster structure, we can say that this state has a structure that the $\Lambda$ particle strongly couples with only one $\alpha$ cluster to form $^5_\Lambda{\rm He}$, while it loosely couples with the other $2\alpha$ clusters. The energy gain of this state from the $0_3^+$ state of $^{12}{\rm C}$, $3.0$ MeV, is very close to the binding energy of $^5_\Lambda{\rm He}$, $3.12$ MeV, and also supports this conclusion.

By comparing the $S^2$-factors between our calculated states in $^{13}_{\Lambda}$C and the ones obtained in Ref.~\cite{Funaki2018plb}, we find that our $0_1^+$, $0_2^+$, $0_4^+$ and $0_5^+$ correspond to the $0_{\rm I}^+$, $0_{\rm II}^+$, $0_{\rm III}^+$ and $0_{\rm IV}^+$ states in Ref.~\cite{Funaki2018plb}.

Finally we discuss the $0_3^+$ state, which has never been discussed before. We find that it has a dominant component of $S^2$-factor in the $^9_{\Lambda}{\rm Be}(0^+_1)+\alpha$ channel, as well as a large component in the $^{12}{\rm C}(0_3^+)+\Lambda$ channel. The RWA in the $^9_{\Lambda}{\rm Be}(0^+_1)+\alpha$ channel shows a higher nodal behaviour, as built on the $0_2^+$ state with the $^9_{\Lambda}{\rm Be}(0^+_1)+\alpha$ structure. The result that the energy gain of this state from the $0_3^+$ state of $^{12}{\rm C}$, $6.7$ MeV, coincides with the binding energy of $^9_\Lambda{\rm Be}$, $6.71$ MeV, also supports the idea that the $\Lambda$ particle strongly couples with the $2\alpha$ clusters to form $^9_\Lambda{\rm Be}$, and higher nodal structure is realized, due to the orthogonalization to the $0_2^+$ state of dominant $^9_\Lambda{\rm Be}+\alpha$ cluster structure.

\begin{acknowledgments}
The authors would like to thank Prof. Emiko Hiyama for useful discussions and support for the code.
This work is supported by the Strategic Priority Research Program of Chinese Academy of Sciences under the Grant NO. XDB34030301, the National Natural Science Foundation of China under the Grant NOs. 12005266, 12075288, 11735003, 11961141012 and Guangdong Major Project of Basic and Applied Basic Research No. 2020B0301030008. It is also
supported by the Youth Innovation Promotion Association CAS. YF acknowledges the financial support by JSPS KAKENHI Grant No. 25400288.

\end{acknowledgments}


\end{document}